\documentclass[fleqn,10pt,onecolumn]{wlscirep}
\usepackage{amsfonts}
\usepackage{txfonts}
\usepackage{amsmath}
\usepackage{float}
\usepackage{graphicx}
\usepackage{dcolumn}
\usepackage{bm}
\usepackage{amssymb}
\usepackage{mathrsfs}
\usepackage{amsmath}
\usepackage{bm}
\usepackage{subfigure}
\usepackage{cases}
\newcommand{\bra}[1]{\left\langle #1\right|}
\newcommand{\ket}[1]{\left|#1\right\rangle}

\title{Simulating Anisotropic quantum Rabi model via frequency modulation}
\author[1]{Gangcheng Wang,*}
\author[1]{Ruoqi Xiao}
\author[1]{H. Z. Shen,$\dag$}
\author[1]{Chunfang Sun}
\author[1]{Kang Xue,$\ddag$}
\affil[1]{Center for Quantum Sciences and School of Physics, Northeast Normal University, Changchun 130024, China}
\affil[*]{Gangcheng Wang (Email: wanggc000@163.com)}
\affil[$\dag$]{H. Z. Shen (Email: shenhz458@nenu.edu.cn)}
\affil[$\ddag$]{Kang Xue (Email: xuekang@nenu.edu.cn)}

\begin{abstract}
Anisotropic quantum Rabi model is a generalization of quantum Rabi model, which allows its rotating and counter-rotating terms to have two different coupling constants. It provides us with a fundamental model to understand various physical features concerning quantum optics, solid-state physics, and mesoscopic physics. In this paper, we propose an experimental feasible scheme to implement anisotropic quantum Rabi model in a circuit quantum electrodynamics system via periodic frequency modulation. An effective Hamiltonian describing the tunable anisotropic quantum Rabi model can be derived from a qubit-resonator coupling system modulated by two periodic driving fields. All effective parameters of the simulated system can be adjusted by tuning the initial phases, the frequencies and the amplitudes of the driving fields. We show that the periodic driving is able to drive a coupled system in dispersive regime to ultrastrong coupling regime, and even deep-strong coupling regime. The derived effective Hamiltonian allows us to obtain pure rotating term and counter-rotating term. Numerical simulation shows that such effective Hamiltonian is valid in ultrastrong coupling regime, and stronger coupling regime. Moreover, our scheme can be generalized to the multi-qubit case. We also give some applications of the simulated system to the Schr\"{o}dinger cat states and quantum gate generalization. The presented proposal will pave a way to further study the stronger anisotropic Rabi model whose coupling strength is far away from ultrastrong coupling and deep-strong coupling regimes in quantum optics.
\end{abstract}

\begin{document}
\flushbottom
\maketitle
\thispagestyle{empty}

\section*{Introduction}
The quantum Rabi model (QRM)~\cite{rabi1936,rabi1937,braak2011} is a fundamental model to describe the light-matter interaction, which has been at the heart of important discoveries of fundamental effects of quantum optics. When the ratio of coupling strength and mode frequency is much smaller than $1$, rotating wave approximation (RWA) is valid and the QRM in this regime can be reduced to the Jaynes-Cummings (JC) model~\cite{jc1963,shore1993}, which has been used to describe the basic interactions in various systems \cite{meekhof1996,leibfried2003,haffner2008,lv2017,blais2004,wallraff2004,blais2007,miller2005,walther2006}. Of particular interest is to implement the QRM in ultra-strong coupling (USC) regime (the coupling strength is comparable to the cavity frequency) \cite{lupascu2015,todorov2010,anappara2009,gunter2009,forndiaz2010,niemczyk2010,federov2010,muravev2011,schwartz2011,scalari2012,geiser2012,goryachev2014,zhang2016,chen2016,solano2003}, and even deep-strong coupling (DSC) regime (the coupling strength exceeds the cavity frequency) \cite{semba2016}, in which RWA is not suitable and the counter-rotating term (CRT) cannot be neglected. This is because various effects induced by CRT appear in these regimes \cite{solano2011,garz2016,wangx2017,Ridolfo2012,Ridolfo2013,Law2013,caox2010,aiq2010,lipb2012,wangx2014,reiter2013,hes2014}. Such tremendous advances in experiments have also motivated various potential applications to quantum information technologies \cite{felicettiprl2014,rossatto2016,kyawprb2015,romero2012,wangym2017}. Although great progresses have been achieved, it is also very challenging to implement such model in USC and DSC regimes experimentally. The quantum simulation proposal provides us with an experimental accessible approach to implement the QRM in USC and DSC regimes, respectively \cite{felicetti2015,lv2016,peder2015,chengxh2018,clerk2018,braum2017,lij2013,ballester2012,mezzacapo2014,langford2017,crespi2012,felicetti2017}.

Recently, a generalized QRM with distinct RT and CRT coupling constants, which has been referred to anisotropic quantum Rabi model (AQRM), is attracting interests \cite{xie2014,tomka2014,shenlt2014,zhangg2015,zhangyy2017,zhangyy2016,yuyx2013}. Due to such interesting characteristics, the AQRM has been utilized to study various theoretical issues, {\rm e.g.}, quantum phase transitions \cite{liu2017,shenlt2017}, quantum state engineering \cite{joshi2016}, quantum fisher information \cite{wangzh2017}, and so on. To date, people have proposed several methods to realize AQRM, which include the natural implementations of AQRM in quantum optics in a cross-electric and magnetic field \cite{xie2014}, electrons in semiconductors with spin-orbit coupling \cite{wangzh2017,schie2003}, and superconducting circuits systems \cite{baksic2014,yangwj2017}. Meanwhile, quantum simulation methods with superconducting circuits \cite{wangym2018} and trapped ions \cite{aedo2018} have also been proposed. The AQRM provides us with a paradigm to understand the light-matter interaction and solid-state system. However, these implementations of AQRM are limited on the tunabilities, which motivate us to develop a frequency modulated method to realize a tunable AQRM in USC or even DSC regimes.

\begin{figure*}
  \centering
  \includegraphics[width=0.8\textwidth]{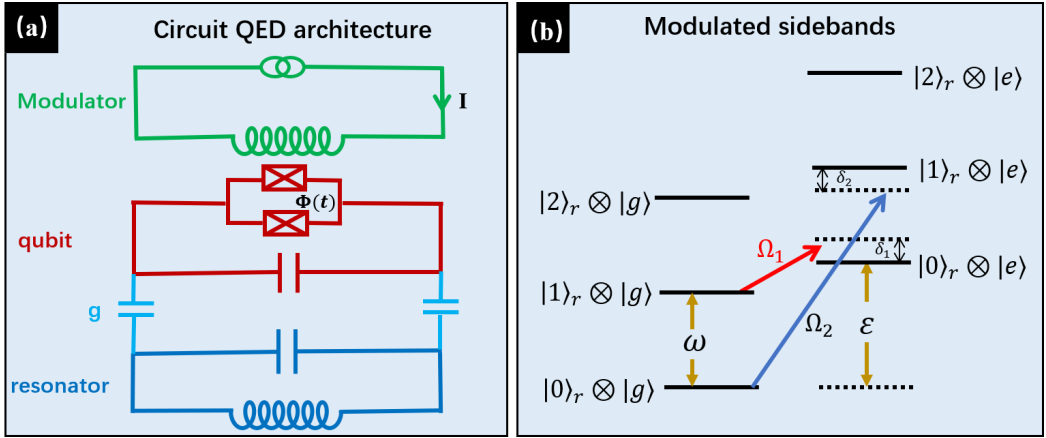}\\
  \caption{(a) The circuit QED architecture of the system: A transmon qubit is capacitively coupled to a LC resonator with frequency $\omega$ \cite{koch2007}. The transmon qubit, which is implemented with split
junctions, can be modulated by the time-dependent flux generated by modulation circuit. The modulation Hamiltonian is shown in Eq. (\ref{eq2c}). (b) The energy level of the modulated qubit-resonator system. Red and blue sidebands detunings of driving fields allow us to tune coupling constants of RT and CRT.}
  \label{fig1}
\end{figure*}

In this paper, we propose an effective method to simulate a tunable AQRM with a qubit coupled to a resonator in dispersive regime, and the transition frequency of the qubit is modulated by two periodic driving fields. The periodic driving have been widely used to modulate quantum systems \cite{strand2013,carlos2014,yanyy2015,liao2016,huangjf2017,silveri2017,basak2018,xuezy2015,xuezy2018_1,xuezy2018_2}. We show that all the parameters in the effective Hamiltonian depend on the external driving fields. The frequencies of qubit and resonator for the simulated system can be adjusted by controlling the frequencies of the driving fields, while the anisotropic coupling coefficients of the RT and CRT are decided by the amplitudes of the driving fields. Our proposal to implement the AQRM has three features: (i) The effective Hamiltonian is controllable, and all the parameters can be tuned by controlling the external driving fields. (ii) We can drive the system from weak-coupling regime to USC regime and even DSC regime by tuning the frequencies and amplitudes of the driving fields. (iii) The ratio of coupling constants of RT and CRT can be controlled in a wide range of parameter space, which makes it possible to study the transitions from JC regime to anti-JC regime.

\section*{The derivation of the effective Hamiltonian}
In this section, we consider a qubit coupled to a harmonic oscillator in dispersive regime, and the qubit is modulated by the periodic driving fields. Such setup can be realized in a variety of different physical contexts, such as trapped ions \cite{meekhof1996,leibfried2003,haffner2008,lv2017}, circuit QED \cite{blais2004,wallraff2004,blais2007}, cavity QED \cite{miller2005,walther2006}, and so on. Here, we adopt a circuit QED setup to illustrate our proposal (the architecture is depicted in Fig.~\ref{fig1}(a)). We consider a tunable transmon qubit, which is comprised of split junctions, is capacitively coupled to a LC resonator. Such split structure allows the qubit to be modulated by the magnetic flux through the pair junctions. The system is described by a time-dependent Hamiltonian as follows (we set $\hbar =1$)
\begin{equation}\label{Eq-II-1}
\hat{H}(t)=\hat{H}_{0}+\hat{H}_{\rm int} +\hat{H}_{d}(t),
\end{equation}
where $\hat{H}_{0}$, $\hat{H}_{\rm int}$ and $\hat{H}_{d}(t)$ are given as follows

\begin{subequations}
\label{eq-II-2}
\begin{align}
\hat{H}_{0} &=\omega \hat{a}^{\dag}\hat{a} + \frac{\varepsilon}{2}\hat{\sigma}_{z},\\
\hat{H}_{\rm int}&=g(\hat{a}+\hat{a}^{\dag})\hat{\sigma}_{x},\\
\hat{H}_{d}&=\sum_{j=1}^{n_{d}}\Omega_{j}\eta_{j}\cos(\Omega_{j}t+\varphi_{j})\hat{\sigma}_{z},\label{eq2c}
\end{align}
\end{subequations}
$\hat{H}(t)=\hat{H}_{0}+\hat{H}_{\mathrm{int}}+\hat{H}_{d}(t)$
where $\varepsilon$ is the transition frequency of the tranmon qubit. $\hat{\sigma}_{\alpha}$ is the $\alpha$-component of the Pauli matrices. $\omega$ is the frequency of the LC resonator. $\hat{a}$ ($\hat{a}^{\dag}$) is the annihilation (creation) operator. $g$ is the coupling constant between the qubit and the bosonic field, and $\hat{H}_{d}(t)$ describes $n_{d}$ periodic driving fields with frequencies $\Omega_{j}$ and normalized amplitudes $\eta_{i}$. In this work, we consider $n_{d}=2$ and the qubit coupled to the resonator in dispersive regime ({\it i.e.}, $|g|\ll |\Delta_{\pm}|$ with $\Delta_{\pm}=\omega \pm \varepsilon$). Without periodic driving, the RT and CRT terms can be ignored in dispersive regime. This is because all terms are fast oscillating terms in the rotating framework. If we choose proper modulation frequencies and amplitudes such that the near resonant physical transitions are remained and far off resonant transitions can be discarded. Moving to the rotating frame defined by the following unitary operator
\begin{eqnarray}
  U_{1}(t) &=& \exp\left(-i\hat{H}_{0}t-i\sum_{j=1}^{2}\eta_{j}\sin(\Omega_{j}t+\varphi_{j})\hat{\sigma}_{z}\right),
\end{eqnarray}
we obtain the transformed Hamiltonian
\begin{small}
\begin{equation}
\begin{aligned}
  \hat{H}'(t) =& U_{1}^{\dag}(t)\hat{H}(t)U_{1}-iU_{1}^{\dag}(t)\left(\partial_{t}U_{1}(t)\right) \\
           =& g\hat{a}\left[\hat{\sigma}_{-}e^{-i\Delta_{+}t}\exp\left(-i\sum_{j=1}^{2}\eta_{j}\sin(\Omega_{j}t+\varphi_{j})\right)+ \hat{\sigma}_{+}e^{-i\Delta_{-}t}\exp\left(i\sum_{j=1}^{2}\eta_{j}\sin(\Omega_{j}t+\varphi_{j})\right)\right]+ {\rm H.c.},
\label{hp}
\end{aligned}
\end{equation}
\end{small}
where $\hat{\sigma}_{\pm}=(\hat{\sigma}_{x}\pm i\hat{\sigma}_{y})/2$. Using the following Jacobi-Anger expansion \cite{colton1998,krosch2006}
 \begin{equation}\label{Iden}
   \exp(2i\eta_{j}\sin(\Omega_{j}t+\varphi_{j}))=\sum_{n=-\infty}^{+\infty}J_{n}(2\eta_{j})\exp[in(\Omega_{j}t+\varphi_{j})],
 \end{equation}
with $J_{n}(x)$ being the $n$-th order Bessel function of the first kind, we obtain
 \begin{equation}
 \label{eq-II-3}
   H'(t)=g\left[\alpha(t)\hat{a}\hat{\sigma}_{+}+\beta(t)\hat{a}\hat{\sigma}_{-}\right]+{\rm H.c.},
 \end{equation}
 where
\begin{small}
\begin{equation}
\begin{aligned}
\alpha(t) & = \sum_{n_{1},n_{2}=-\infty}^{+\infty}J_{n_{1}}(2\eta_{1})J_{n_{2}}(2\eta_{2})e^{i(n_{1}\varphi_{1}+n_{2}\varphi_{2})}e^{i\Omega_{-}(n_{1},n_{2})t},\\
\beta(t)  & = \sum_{n_{1},n_{2}=-\infty}^{+\infty}J_{n_{1}}(2\eta_{1})J_{n_{2}}(2\eta_{2})e^{-i(n_{1}\varphi_{1}+n_{2}\varphi_{2})}e^{-i\Omega_{+}(n_{1},n_{2})t}.
\label{eq-II-4}
\end{aligned}
\end{equation}
\end{small}
Here, $\Omega_{\pm}(n_{1},n_{2})=\Delta_{\pm}+ n_{1}\Omega_{1}+ n_{2}\Omega_{2}$. According to the RWA, only slowly varying terms appearing in $\alpha(t)$ and $\beta(t)$ will dominate the dynamics. We should choose the suitable driving frequencies to obtain the rotating and counter-rotating interaction terms. We assume there is a small detuning $\delta_{1}$ ($\delta_{2}$) between $\Omega_{1}$ ($\Omega_{2}$) and the red (blue) sideband, and the definition of the detunings read
 \begin{equation}
\label{eq-II-4-1}
\delta_{1}=\Omega_{1}-\Delta_{-},\quad \delta_{2}=\Delta_{+}-\Omega_{2}.
\end{equation}
The energy levels of the modulated system are shown in Fig.~\ref{fig1}(b). Considering small detunings (\emph{i.e.} $|\delta_{i}|\ll |\Delta_{\pm}|$) and dispersive coupling regime (\emph{i.e.} $|g|\ll |\Delta_{\pm}|$), one can check that the RT and the CRT will contribute to the dynamics only for lowest oscillating frequencies $\Omega_{-}(-1,0)=-\delta_{1}$ and $\Omega_{+}(0,-1)=\delta_{2}$, respectively. When the oscillating frequencies are much larger than the effect couplings, {\rm i.e.}, $|\Omega_{+}(m_{1},m_{2})|\gg |g J_{m_{1}}(2\eta_{1})J_{m_{2}}(2\eta_{2})|$ with $(m_{1},m_{2})\neq (0,-1)$ and $|\Omega_{-}(q_{1},q_{2})|\gg |g J_{q_{1}}(2\eta_{1})J_{q_{2}}(2\eta_{2})|$ with $(q_{1},q_{2})\neq (-1,0)$, one may safely neglect these fast oscillating terms in Eq. (\ref{eq-II-3}). Then the dominant terms in Eq.~(\ref{eq-II-4}) are $\alpha(t) \approx -J_{1}(2\eta_{1})J_{0}(2\eta_{2})e^{-i\varphi_{1}}e^{-i\delta_{1}t}$ and $\beta(t) \approx -J_{0}(2\eta_{1})J_{1}(2\eta_{2})e^{i\varphi_{2}}e^{-i\delta_{2}t}$, where we have used the relation $J_{-n}(x)=(-1)^{n}J_{n}(x)$ for integer $n$. Then these approximations lead to the following near resonant time-dependent Hamiltonian
\begin{small}
\begin{equation}
\begin{aligned}
  \hat{H}'(t) &\approx & \left(\tilde{g}_{r}\hat{a}\hat{\sigma}^{+}e^{-i(\delta_{1}t+\varphi_{1})} +  \tilde{g}_{cr}\hat{a}^{\dag}\hat{\sigma}^{+}e^{i(\delta_{2}t-\varphi_{2})}\right)+{\rm H.c.},
  \label{eq-II-5}
\end{aligned}
\end{equation}
\end{small}
where the effective coupling strengths of RT and CRT are
 \begin{equation}
\label{eq-II-6}
\tilde{g}_{r}=-gJ_{1}(2\eta_{1})J_{0}(2\eta_{2}),\quad \tilde{g}_{cr}=-gJ_{0}(2\eta_{1})J_{1}(2\eta_{2}).
\end{equation}

The Hamiltonian in Eq. (\ref{eq-II-5}) is the so-called AQRM in interaction picture with effective resonator frequency $\tilde{\omega}=(\delta_{1}+\delta_{2})/2$ and qubit transition frequency $\tilde{\varepsilon}=(\delta_{2}-\delta_{1})/2$.
Defining the new rotating framework associated with the time-dependent unitary operator
\begin{eqnarray}
U_{2}(t)=\exp(i\tilde{\omega}\hat{a}^{\dag}\hat{a}t+i\frac{\tilde{\varepsilon}}{2}\hat{\sigma}_{z}t),
\end{eqnarray}
we obtain the effective Hamiltonian with anisotropic coupling strengths for RT and CRT
  \begin{eqnarray}
  \label{Heff}
  \hat{H}_{\rm eff} &=&  \tilde{\omega}\hat{a}^{\dag}\hat{a}+\frac{\tilde{\varepsilon}}{2}\hat{\sigma}_{z}+ \tilde{g}_{r}\left(\hat{a}\hat{\sigma}_{+}+\hat{a}^{\dag}\hat{\sigma}_{-}\right)+\tilde{g}_{cr}\left(\hat{a}\hat{\sigma}_{-}e^{i\theta}+\hat{a}^{\dag}\hat{\sigma}_{+}e^{-i\theta}\right),
\end{eqnarray}
where we have set $\varphi_{1} = 0$ and $\varphi_{2} = \theta$. The anisotropic parameter $\lambda$ is the ratio of RT and CRT coupling strengths ({\it i.e.}, $\lambda=\tilde{g}_{cr}/\tilde{g}_{r}$). Thus we obtain a controllable AQRM. Below we analyze the parameters in our scheme. In our circuit QED setup, we consider the following realistic parameters \cite{goerz2017,hofh2009}: the transition frequency of the transmon qubit is $\varepsilon = 2\pi\times 5.4~{\rm GHz}$ with the decay rate $\kappa = 2\pi\times 0.05~{\rm MHz}$, the resonator frequency is $\omega=2\pi\times 2.2~{\rm GHz}$ with the loss rate $\gamma = 2\pi\times 0.012~{\rm MHz}$, and the coupling strength of the resonator and qubit is $g = 2\pi\times 70~{\rm MHz}$. We can check that the dispersive condition ({\rm i.e.}, $|g|\ll |\Delta_{\pm}|$) is fulfilled. The frequency modulation can be implemented by applying proper biasing magnetic fluxes. The modulation parameters $\Omega_{i}$, $\eta_{i}$ and $\varphi_{i}$ can be chosen on demand by tuning the modulation fields. In circuit QED setups, the modulation frequency and modulation amplitude range from hundreds of megahertz to several gigahertz. It is reasonable to set the modulation amplitude $\eta_{i}\Omega_{i}$ ranges from $0$ to $2\pi\times10~{\rm GHz}$ \cite{lupascu2015}. The detunings $\delta_{i}$ can be tuned from $0$ to hundreds of megahertz to fulfill the condition $|\delta_{i}|\ll |\Delta_{\pm}|$.

\begin{table}
\centering
\caption{The system parameters are listed.}
\label{tab-I}
\begin{tabular}{ccccc}
\hline
\hline
$\varepsilon/2\pi$ & $\omega/2\pi$ & $g/2\pi$  & $\gamma/2\pi$ & $\kappa/2\pi$ \\
\hline
5.4~GHz & 2.2~GHz & 70~MHz & 12~KHz & 50~KHz\\
\hline
\hline
\end{tabular}
\end{table}

\section*{The simulation of QRM and AQRM in USC and DSC regimes}
\label{sec-III}
To assess the robustness of our proposal in circuit QED system, we should consider the dissipation effects in the following discussions~\cite{goerz2017}. Considering the zero-temperature Markovian environments and large driven frequencies $\Omega_{j}$, the master equation governing the evolution of the system can be derived as follows \cite{clerk2018}

\begin{equation}\label{master_eq}
  \dot{\rho} = -i[\hat{H}(t),\rho]+\mathcal{L}_{q}[\rho]+\mathcal{L}_{r}[\rho],
\end{equation}
where $\mathcal{L}_{q}[\rho]=\frac{\kappa}{2}(2\hat{\sigma}_{-}\rho\hat{\sigma}_{+}-\rho\hat{\sigma}_{+}\hat{\sigma}_{-}-\hat{\sigma}_{+}\hat{\sigma}_{-}\rho)$ and $\mathcal{L}_{r}=\frac{\gamma}{2}(2\hat{a}\rho \hat{a}^{\dag}-\rho \hat{a}^{\dag}\hat{a}-\hat{a}^{\dag}\hat{a}\rho)$ are the standard Lindblad super-operators describing the losses of the system. To obtain the master equation in the framework of effective Hamiltonian, we set $U(t)=U_{2}(t)U_{1}(t)$. Let $\tilde{\rho}(t)$ be the density matrix in the same framework with effective Hamiltonian. Inserting $\rho(t)=U(t)\tilde{\rho}(t)U^{\dag}(t)$ to the master equation (\ref{master_eq}), we obtain the following master equation
\begin{equation}\label{master_eq_new}
  \dot{\tilde{\rho}} = -i[\hat{\tilde{H}}(t),\tilde{\rho}]+\mathcal{L}_{q}[\tilde{\rho}]+\mathcal{L}_{r}[\tilde{\rho}],
\end{equation}
where $\hat{\tilde{H}}(t)=U^{\dag}(t)\hat{H}(t)U(t)-iU^{\dag}(t)\left(\partial_{t}U(t)\right)$ is the total system Hamiltonian in the new rotating framework. We show that the Hamiltonian $\hat{H}_{\rm eff}$ in Eq.~(\ref{Heff}) is the approximation of $\hat{\tilde{H}}(t)$ under RWA. Here we consider the initial phase difference of the driving fields is $\theta=0$. The parameters $\eta_{1,2}$ and the detuning of first sideband $\delta_{1,2}$ are tunable parameters. Such tunable parameters determine the parameters in the simulated system in Eq.~(\ref{Heff}). To verify the validity of the effective Hamiltonian in Eq.~(\ref{Heff}), we should study the fidelity of the evolution state. Let $\ket{\tilde{\psi}(0)}$ be an initial state in the new framework and the corresponding initial density matrix is $\tilde{\rho}(0)=\ket{\tilde{\psi}(0)}\bra{\tilde{\psi}(0)}$. Substituting $\tilde{\rho}(0)$ into Eq.~(\ref{master_eq_new}), we obtain the evolution density matrix $\tilde{\rho}(t)$. The ideal case can be obtained by solving the Sch\"{o}rdinger equation governed by the effective Hamiltonian (\ref{Heff}). We denote the ideal evolution state governed by the effective Hamiltonian (\ref{Heff}) with $\ket{\tilde{\psi}(t)}$. Then the fidelity of the evolution state reads $F(t)=\left|\bra{\tilde{\psi}(t)}\tilde{\rho}(t)\ket{\tilde{\psi}(t)}\right|$.

\subsection*{The simulation of QRM}
In this subsection, we will show the performance of the simulated QRM. To obtain equal effective RT and CRT coupling strengths ({\rm i.e.}, $\lambda=1$), we need to adjust the normalized amplitude $\eta_{i}$. A simple case is $\eta_{1}=\eta_{2}=\eta$. Then the simulated coupling strength $\tilde{g}_{r}=\tilde{g}_{cr}$ and we denote the simulated coupling strength with $\tilde{g}=-gJ_{0}(2\eta)J_{1}(2\eta)$. Assuming $\theta = 0$, we can obtain the following tunable QRM
\begin{eqnarray}
  \label{QRM}
  H_{\rm QRM} &=&  \tilde{\omega}\hat{a}^{\dag}\hat{a}+\frac{\tilde{\varepsilon}}{2}\hat{\sigma}_{z}+\tilde{g}\left(\hat{a}^{\dag}+\hat{a}\right)\hat{\sigma}_{x}.
\end{eqnarray}
 The effective frequencies of resonator and qubit are determined by the detunings $\delta_{i}$. One can tuning the ratios of modulation amplitudes and frequencies to obtain different relative coupling strength.

\begin{figure*}[t]
  \centering
  \includegraphics[width=0.8\textwidth]{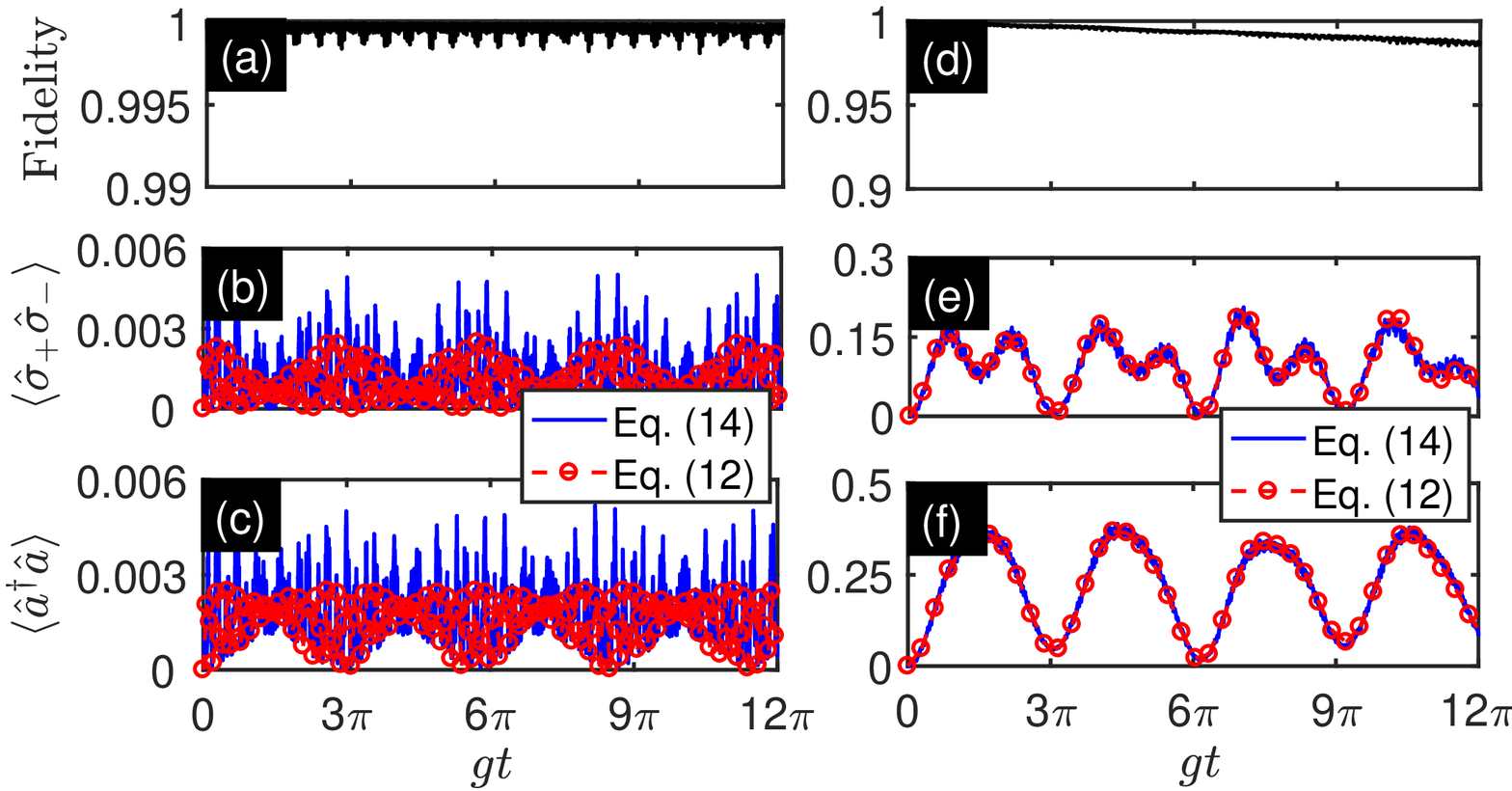}\\
  \caption{The fidelity and dynamics of the simulated QRM with effective coupling ratio $|\tilde{g}/\tilde{\omega}|=0.05$ (Figures (a),(b),(c)) and $|\tilde{g}/\tilde{\omega}|=0.5$ (Figures (d), (e), (f)) as a functions of evolution time. Figures (a) and (d) show the fidelity as a function of evolution time governed by the master equation in Eq.~(\ref{master_eq_new}) and the simulated Hamiltonian given in Eq.~(\ref{Heff}). Figures (b) and (e) show the qubit excitation number $\langle \hat{\sigma}_{+}\hat{\sigma}_{-}\rangle$ as a function of evolution time. Figures (c) and (f) show the excitation number of the resonator $\langle \hat{a}^{\dag}\hat{a}\rangle$ as a function of evolution time. The dynamics is governed by the master equation in Eq.~(\ref{master_eq_new}) (solid blue line) and the simulated Hamiltonian given in Eq.~(\ref{Heff}) (red dashed line with circles). The red sideband modulation parameters are chosen as $\Omega_{1}=2\pi\times 3.2~{\rm GHz}$, and $\eta_{1}\Omega_{1}=2\pi\times 2.296~{\rm GHz}$. The blue sideband modulation parameters are chosen as follows: $\Omega_{2}= 2\pi\times 6.759~{\rm GHz}$, and $\eta_{2}\Omega_{2}=2\pi\times 4.849~{\rm GHz}$ for figures (a), (b), (c) and $\Omega_{2}= 2\pi\times 7.516~{\rm GHz}$, and $\eta_{2}\Omega_{2}=2\pi\times 5.392~{\rm GHz}$ for figures (d), (e), (f). The initial state is prepared on the state $\ket{\tilde{\psi}(0)}=\ket{0}_{r}\otimes\ket{g}$. The other parameters are listed in Table~\ref{tab-I}.}
  \label{fig2}
\end{figure*}

\begin{figure*}[ht]
  \centering
  \includegraphics[width=0.8\textwidth]{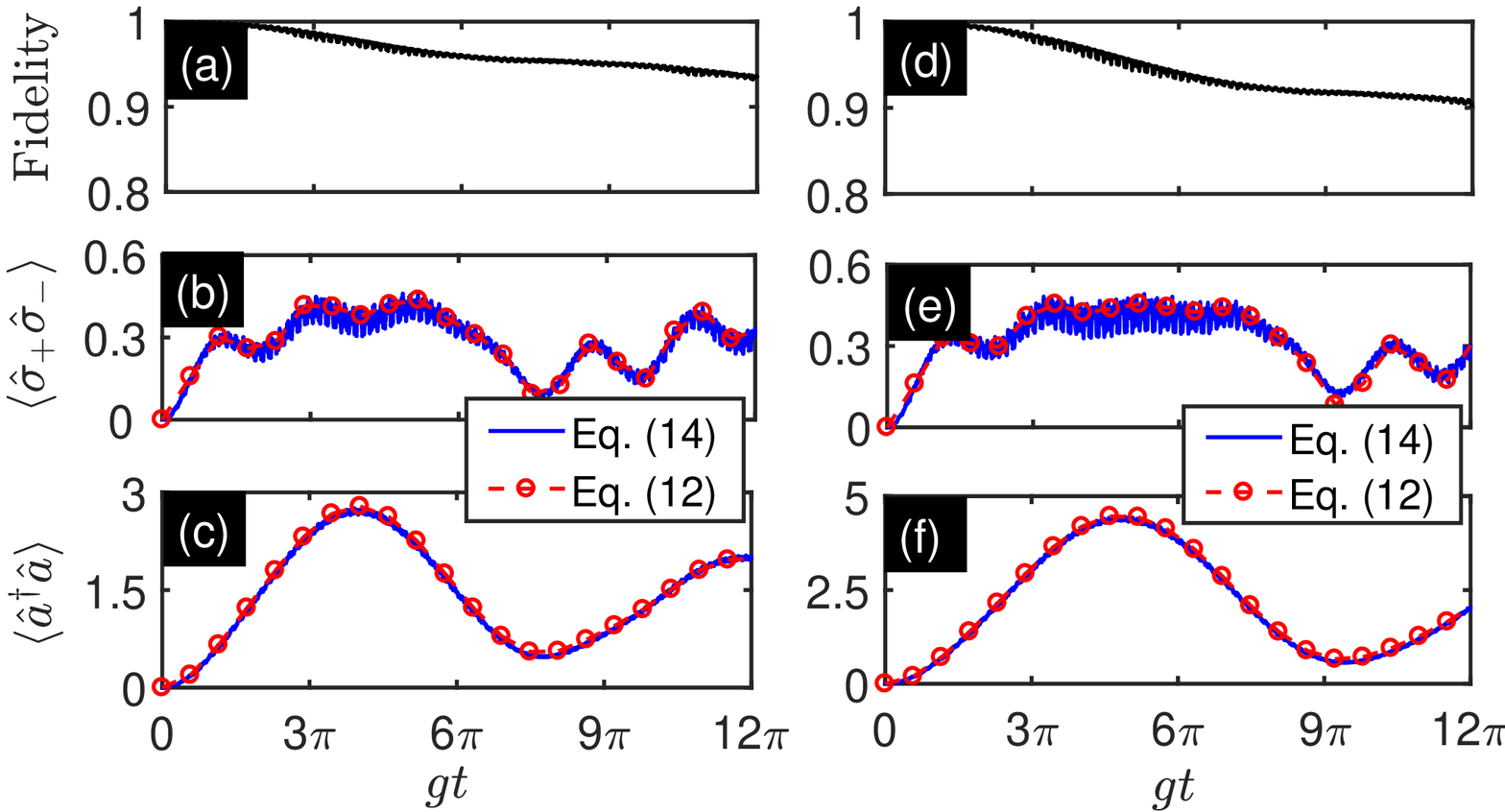}\\
  \caption{The fidelity and dynamics of simulated QRM with effective coupling ratio $|\tilde{g}/\tilde{\omega}|=1$ (Figures (a),(b),(c)) and $|\tilde{g}/\tilde{\omega}|=1.2$ (Figures (d), (e), (f)) as functions of evolution time. Figures (a) and (d) show the fidelity as a function of evolution time governed by the master equation in Eq.~(\ref{master_eq_new}) and the simulated Hamiltonian given in Eq.~(\ref{Heff}). Figures (b) and (e) show the qubit excitation number $\langle \hat{\sigma}_{+}\hat{\sigma}_{-}\rangle$ as a function of evolution time. Figures (c) and (f) show the excitation number of the resonator $\langle a^{\dag}a\rangle$ as function of evolution time. The dynamics is governed by the master equation in Eq.~(\ref{master_eq_new}) (blue solid line) and the simulated Hamiltonian given in Eq.~(\ref{Heff}) (red dashed line with circles). The red sideband modulation parameters are chosen as $\Omega_{1}=2\pi\times 3.2~{\rm GHz}$, and $\eta_{1}\Omega_{1}=2\pi\times 2.296~{\rm GHz}$. The blue sideband modulation parameters are chosen as follows: $\Omega_{2}= 2\pi\times 7.558~{\rm GHz}$, and $\eta_{2}\Omega_{2}=2\pi\times 5.422~{\rm GHz}$ for figures (a), (b), (c) and $\Omega_{2}= 2\pi\times 7.565~{\rm GHz}$, and $\eta_{2}\Omega_{2}=2\pi\times 5.427~{\rm GHz}$ for figures (d), (e), (f). The initial state is prepared on the state $\ket{\tilde{\psi}(0)}=\ket{0}_{r}\otimes\ket{g}$. The other parameters are listed in Table~\ref{tab-I}.}
  \label{fig3}
\end{figure*}

 In Figs. \ref{fig2} and \ref{fig3}, we show the fidelity and dynamics under following four sets parameters: \ref{fig2}(a), \ref{fig2}(b) and \ref{fig2}(c) $\Omega_{2}= 2\pi\times 6.759~{\rm GHz}$, and $\eta_{2}\Omega_{2}=2\pi\times 4.849~{\rm GHz}$; \ref{fig2}(d), \ref{fig2}(e) and \ref{fig2}(f) $\Omega_{2}= 2\pi\times 7.516~{\rm GHz}$, and $\eta_{2}\Omega_{2}=2\pi\times 5.392~{\rm GHz}$; \ref{fig3}(a), \ref{fig3}(b) and \ref{fig3}(c) $\Omega_{2}= 2\pi\times 7.558~{\rm GHz}$, and $\eta_{2}\Omega_{2}=2\pi\times 5.422~{\rm GHz}$; \ref{fig3}(d), \ref{fig3}(e) and \ref{fig3}(f) $\Omega_{2}= 2\pi\times 7.565~{\rm GHz}$, and $\eta_{2}\Omega_{2}=2\pi\times 5.427~{\rm GHz}$. The red sideband modulation parameters are chosen as $\Omega_{1}=2\pi\times 3.2~{\rm GHz}$, and $\eta_{1}\Omega_{1}=2\pi\times 2.296~{\rm GHz}$. These sets parameters imply the normalized modulation amplitudes $\eta=0.7173$. One also can lead to resonant red sideband ({\rm i.e.}, $\delta_{1}=0$) and the detuned blue sideband, and the corresponding detunings read $\delta_{2}=2\pi\times 840.7~{\rm MHz}$, $2\pi\times 84.07~{\rm MHz}$, $2\pi\times 42.03~{\rm MHz}$ and $2\pi\times 35.03~{\rm MHz}$. These sets parameters correspond to the four relative coupling strengths $|\tilde{g}/\tilde{\omega}|=0.05$ (Figures~\ref{fig2}(a), \ref{fig2}(b), \ref{fig2}(c)), $|\tilde{g}/\tilde{\omega}|=0.5$ (Figures~\ref{fig2}(d), \ref{fig2}(e), \ref{fig2}(f)), $|\tilde{g}/\tilde{\omega}|=1$ (Figures~\ref{fig3}(a), \ref{fig3}(b), \ref{fig3}(c)) and $|\tilde{g}/\tilde{\omega}|=1.2$ (Figures~\ref{fig3}(d), \ref{fig3}(e), \ref{fig3}(f)). In the numerical simulation, we take $\ket{\tilde{\psi}(0)}=\ket{0}_{r}\otimes\ket{g}$ as initial state. Figures~\ref{fig2}(a) and \ref{fig2}(d) show the fidelity as a function of evolution time governed by the master equation in Eq.~(\ref{master_eq_new}) and the simulated Hamiltonian given in Eq.~(\ref{Heff}). Figures~\ref{fig2}(b) and \ref{fig2}(e) show the qubit excitation number $\langle \hat{\sigma}_{+}\hat{\sigma}_{-}\rangle$ as a function of evolution time. Figures~\ref{fig2}(c) and \ref{fig2}(f) show the excitation number of the resonator $\langle \hat{a}^{\dag}\hat{a}\rangle$ as a function of evolution time. The dynamics is governed by the master equation in Eq.~(\ref{master_eq_new}) (blue solid line) and the simulated Hamiltonian given in Eq.~(\ref{Heff}) (red dashed line with circles). In the case of $|\tilde{g}/\tilde{\omega}|=0.05$, the RWA is valid and the dynamics of qubit and the resonator are dominated by RT. The effects of CRT are very weak, and we can apply RWA safely. In the case of $|\tilde{g}/\tilde{\omega}|=0.5$, the RWA is not valid and the effects of CRT cannot be ignored. The qubit and resonator can be excited simultaneously. The Fig.~\ref{fig3} shows the fidelity and dynamics when $|\tilde{g}/\tilde{\omega}|=1$ (Figures~\ref{fig3}(a), \ref{fig3}(b), \ref{fig3}(c)) and $|\tilde{g}/\tilde{\omega}|=1.2$ (Figures~\ref{fig3}(d), \ref{fig3}(e), \ref{fig3}(f)). In these cases, the relative effective coupling strength reaches $1$ and even exceeds $1$. The CRT plays an important role in USC and DSC regimes. The exist of CRT makes the total excitation number operator $\hat{N}=\hat{a}^{\dag}\hat{a}+\hat{\sigma}_{+}\hat{\sigma}_{-}$ not a conserved quantity. The excitations of qubit and resonator can be excited from the vacuum. The Figures~\ref{fig3}(d), \ref{fig3}(e), \ref{fig3}(f) show the fidelity and dynamic when $|\tilde{g}/\tilde{\omega}|=1.2$. In this case, DSC regime is reached.

\subsection*{The simulation of JC model and anti-JC model}
\begin{figure*}[ht]
  \centering
  \includegraphics[width=0.8\textwidth]{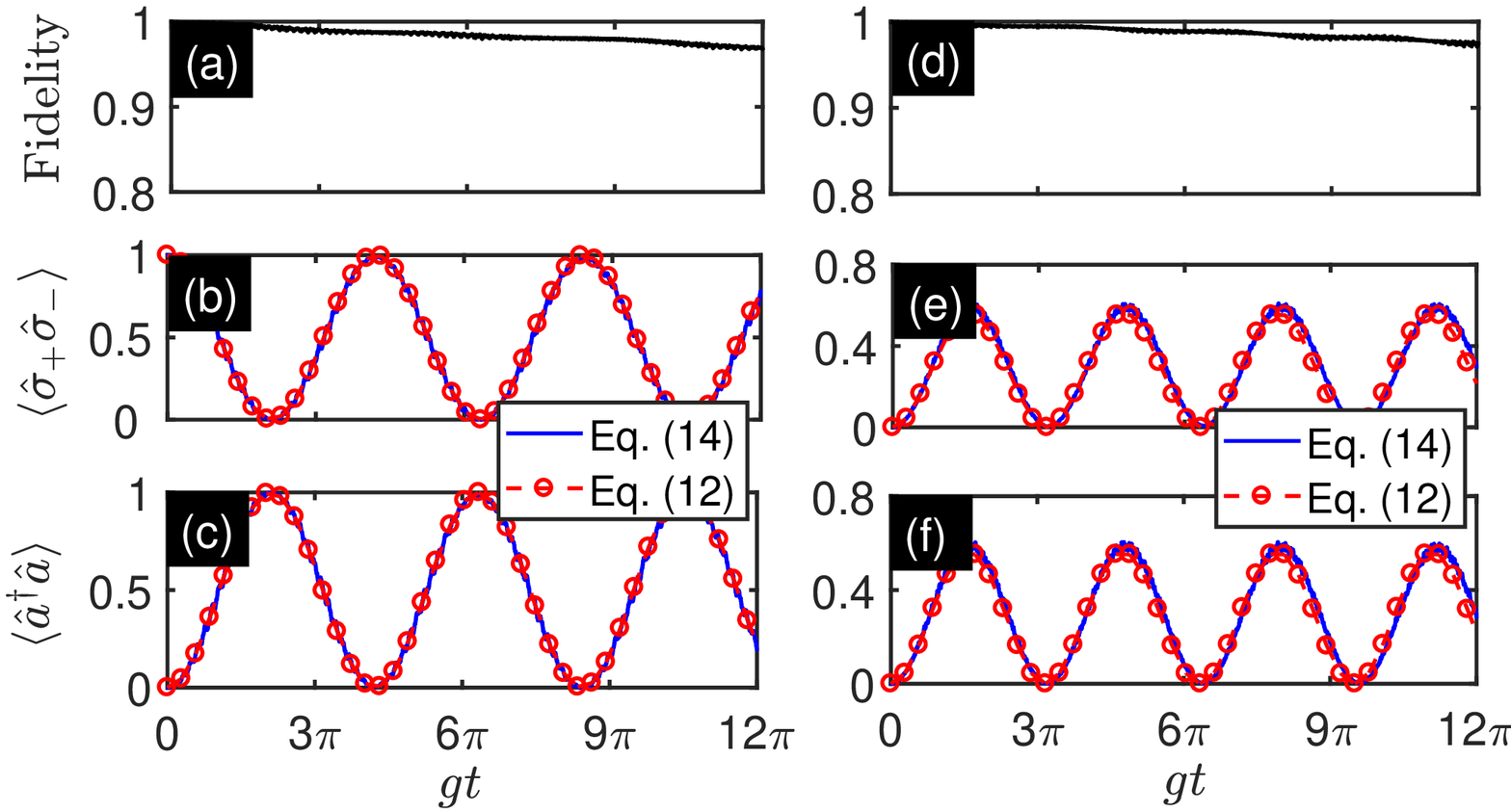}\\
  \caption{The fidelity and dynamics of simulated QRM with effective coupling ratio $|\tilde{g}_{r}/\tilde{\omega}|=1.137$ (Figures (a),(b),(c)) and $|\tilde{g}_{cr}/\tilde{\omega}|=1.137$ (Figures (d), (e), (f)) as functions of evolution time. Figures (a) and (d) show the fidelity as a function of evolution time are governed by the master equation in Eq.~(\ref{master_eq_new}) and the simulated Hamiltonian given in Eq.~(\ref{Heff}). Figures (b) and (e) show the qubit excitation number $\langle \hat{\sigma}_{+}\hat{\sigma}_{-}\rangle$ as a function of evolution time. Figures (c) and (f) show the excitation number of the resonator $\langle \hat{a}^{\dag}\hat{a}\rangle$ as a function of evolution time. The dynamics is governed by the master equation in Eq.~(\ref{master_eq_new}) (blue solid line) and the simulated Hamiltonian is given in Eq.~(\ref{Heff}) (red dashed line with circles). The parameters are taken as follows: $\Omega_{1}=2\pi\times 3.2~{\rm GHz}$, $\eta_{1}\Omega_{1}=2\pi\times 3.848~{\rm GHz}$, $\Omega_{2}=2\pi\times 7.565~{\rm GHz}$, and $\eta_{2}\Omega_{2}=2\pi\times 5.427~{\rm GHz}$, $\varphi_{1}=\varphi_{2}=0$ for figures (a), (b), (c) and $\Omega_{1}=2\pi\times 3.2~{\rm GHz}$, $\eta_{1}\Omega_{1}=2\pi\times 2.296~{\rm GHz}$, $\Omega_{2}=2\pi\times 7.565~{\rm GHz}$, and $\eta_{2}\Omega_{2}=2\pi\times 9.096~{\rm GHz}$, $\varphi_{1}=\varphi_{2}=0$ for figures (d), (e), (f). The other parameters are listed in Table~\ref{tab-I}.}
  \label{fig4}
\end{figure*}

 In this subsection, we will show how to obtain the JC model and anti-JC model by tuning the driving parameters to suppress the CRT or RT, respectively. To obtain the JC model, we chosen the modulation parameters as follows: $\Omega_{1}=2\pi\times 3.2~{\rm GHz}$, $\eta_{1}\Omega_{1}=2\pi\times 3.848~{\rm GHz}$, $\Omega_{2}=2\pi\times 7.565~{\rm GHz}$, and $\eta_{2}\Omega_{2}=2\pi\times 5.427~{\rm GHz}$, $\varphi_{1}=\varphi_{2}=0$. The other parameters are listed in Table \ref{tab-I}. These modulation parameters imply $\eta_{1}=1.2024$, $\eta_{2}=0.7173$, $\delta_{1}=0$, $\delta_{2}=2\pi\times 35.03$ {\rm MHz} and $\theta=0$. One can check that $\tilde{g}_{cr}=0$ and the relative coupling strength $|\tilde{g}_{r}/\tilde{\omega}|=1.137$. In this case, the rotating term is suppressed to zero and the effective Hamiltonian reduced to the following the JC model in DSC regime
 \begin{eqnarray}
  \label{JC}
  H_{\rm JC} &=&  \tilde{\omega}a^{\dag}a+\frac{\tilde{\varepsilon}}{2}\hat{\sigma}_{z}+\tilde{g}_{r}\left(a\hat{\sigma}_{+}+a^{\dag}\hat{\sigma}_{-}\right).
\end{eqnarray}
Taking the initial state $\ket{\tilde{\psi}(0)}=\ket{0}_{r}\otimes\ket{e}$, we obtain the fidelity and dynamics of the evolution state governed by the master equation in Eq.~(\ref{master_eq_new}) and the simulated Hamiltonian given in Eq.~(\ref{JC}), which are shown in figures \ref{fig4}(a), \ref{fig4}(b) and \ref{fig4}(c). The results show that the numerical simulation agrees well with the exact dynamics. It also shows that there exists the Rabi oscillation between states $\ket{0}_{r}\otimes\ket{e}$ and $\ket{1}_{r}\otimes\ket{g}$ with period $\pi/|\tilde{g}_{r}|$. For the case $\delta_{1}\neq 0$ and the initial state $\ket{\tilde{\psi}(0)}=\ket{0}_{r}\otimes\ket{e}$, the period of the Rabi oscillation is $2\pi/\sqrt{4\tilde{g}_{r}^{2}+\delta_{1}^{2}}$.

 To obtain the anti-JC model, we set $\Omega_{1}=2\pi\times 3.2~{\rm GHz}$, $\eta_{1}\Omega_{1}=2\pi\times 2.296~{\rm GHz}$, $\Omega_{2}=2\pi\times 7.565~{\rm GHz}$, and $\eta_{2}\Omega_{2}=2\pi\times 9.096~{\rm GHz}$, $\varphi_{1}=\varphi_{2}=0$. The other parameters are listed in Table \ref{tab-I}. These modulation parameters imply $\eta_{1}=0.7173$, $\eta_{2}=1.2024$, $\delta_{1}=0$, $\delta_{2}=2\pi\times 35.03$ {\rm MHz} and $\theta=0$. In this case, we can check that $\tilde{g}_{r}=0$ and the relative coupling strength $|\tilde{g}_{cr}/\tilde{\omega}|=1.137$. The effective Hamiltonian is reduced to the following anti-JC model in DSC regime
\begin{eqnarray}
  \label{AJC}
  H_{\rm AJC} &=&  \tilde{\omega}\hat{a}^{\dag}\hat{a}+\frac{\tilde{\varepsilon}}{2}\hat{\sigma}_{z}+\tilde{g}_{cr}\left(\hat{a}\hat{\sigma}_{-}+\hat{a}^{\dag}\hat{\sigma}_{+}\right).
\end{eqnarray}
In the anti-JC model, the rotating term is suppressed to zero and only the CRT remains. We can check the validity and dynamics of the effective Hamiltonian. Let $\ket{\tilde{\psi}(0)}=\ket{0}_{r}\otimes\ket{g}$ be the initial state. The fidelity and dynamics of the evolution state governed by the master equation in Eq.~(\ref{master_eq_new}) and the simulated Hamiltonian given in Eq.~(\ref{AJC}) are shown in figures \ref{fig4}(d), \ref{fig4}(d), \ref{fig4}(f). The Fig.~\ref{fig4}(d) shows that the numerical simulation agrees well with the exact dynamics. The Figures~\ref{fig4}(e) and \ref{fig4}(f) show that excitation number of the resonator and qubit possesses the same behavior. It also shows that there exists the Rabi oscillation between states $\ket{0}_{r}\otimes\ket{g}$ and $\ket{1}_{r}\otimes\ket{e}$ with period $2\pi/\sqrt{4\tilde{g}_{cr}^{2}+\delta_{2}^{2}}$. For the case $\delta_{2}=0$, the period of the Rabi oscillation is $\pi/|\tilde{g}_{cr}|$. Such behavior is induced by pure effect of CRT and have been studied in Ref.~\cite{Garziano2015}.

\subsection*{The simulation of degenerate AQRM}
In this subsection, we will simulate the dynamics of the AQRM. For simplify, we choose the modulation parameters are as follows: $\Omega_{1}=2\pi\times 3.2~{\rm GHz}$, $\eta_{1}\Omega_{1}=2\pi\times 2.296~{\rm GHz}$, $\Omega_{2}=2\pi\times 7.6~{\rm GHz}$, $\varphi_{1}=\varphi_{2}=0$, and the blue sideband modulation amplitude ranges from $0$ to $2\pi\times 9.138~{\rm GHz}$. The other parameters are given in Table \ref{tab-I}. Then we can obtain $\delta_{1}=\delta_{2}=0$, $\eta_{1}=0.7173$ and $\theta=0$. The normalized amplitude of blue sideband ranges from $0$ to $1.2024$. In this case, only interaction terms remain and the effective Hamiltonian reduces to the following degenerate AQRM

\begin{eqnarray}
  \label{H-DAQRM}
  \hat{H}_{\rm DAQRM} &=&  \tilde{g}_{r}\left(\hat{a}\hat{\sigma}_{+}+\hat{a}^{\dag}\hat{\sigma}_{-}\right)+\tilde{g}_{cr}\left(\hat{a}\hat{\sigma}_{-}+\hat{a}^{\dag}\hat{\sigma}_{+}\right).
\end{eqnarray}

\begin{figure*}[ht]
  \centering
  \includegraphics[width=0.8\textwidth]{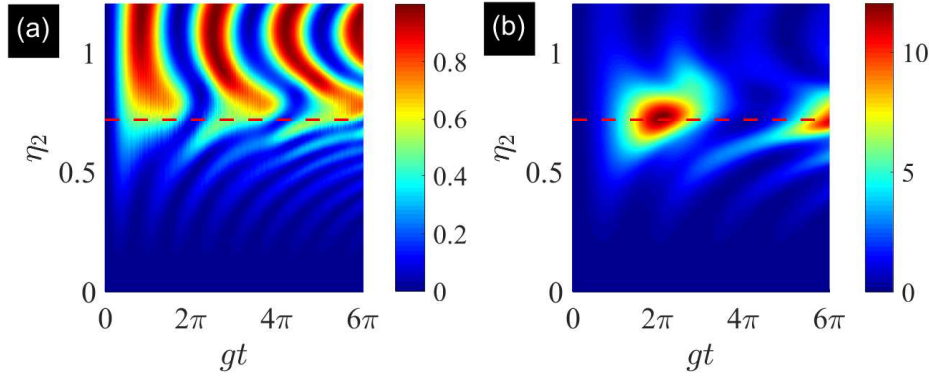}\\
    \caption{The dynamics of simulated degenerate AQRM as a function of evolution time and $\eta_{2}$. (a) shows the excitation of qubit $\langle\hat{\sigma}_{+}\hat{\sigma}_{-}\rangle$ as a function of evolution time and $\eta_2$. (b) shows the excitation of resonator $\langle\hat{a}^{\dagger}\hat{a}\rangle$ as a function of evolution time and $\eta_2$.The parameters are taken as follows: $\Omega_{1}=2\pi\times 3.2~{\rm GHz}$, $\eta_{1}\Omega_{1}=2\pi\times 2.296~{\rm GHz}$, $\Omega_{2}=2\pi\times 7.6~{\rm GHz}$, $\varphi_{1}=\varphi_{2}=0$, and the blue sideband modulation amplitude ranges from $0$ to $2\pi\times 9.138~{\rm GHz}$ ({\rm i.e.}, $\eta_{2}$ ranges from $0$ to $1.2024$). The other parameters are given in Table \ref{tab-I}. The initial state is chosen as $\ket{\tilde{\psi}(0)}=\ket{0}_{r}\otimes\ket{g}$. The red dashed line is plotted for $\eta_{2}=0.7173$. The evolution states are governed by the master equation~(\ref{master_eq_new}).}
  \label{fig5}
\end{figure*}

We check that the simulated Hamiltonian varies from JC model to anti-JC model by tuning the normalized amplitude $\eta_{2}$. Let $\ket{\tilde{\psi}(0)}=\ket{0}_{r}\otimes\ket{g}$ be the initial state. We can obtain the dynamics of the evolution states governed by the master equation in Eq.~(\ref{master_eq}). The excitations of qubit and resonator as a function of evolution time and $\eta_2$ are shown in Fig.~\ref{fig5}. The Fig.~\ref{fig5}(a) shows the excitation of qubit $\langle\hat{\sigma}_{+}\hat{\sigma}_{-}\rangle$ as a function of evolution time and $\eta_2$. When $\eta_{2}\ll 1$, we can check that $\tilde{g}_{cr}$ approaches to zero and rotating term dominates the dynamics. The qubit and resonator are not excited in the evolution process. If we increase the normalized amplitude $\eta_{2}$, the effects of the CRT emerge. In this regime, the qubit and resonator are excited in the evolution process. When $\eta_{2} = 0.7173$ (red dashed line), the ration of the RT and CRT approaches to $1$. In this regime, the RT and CRT dominate the dynamics of the evolution. The Fig.~\ref{fig5}(b) shows the excitation number of resonator $\langle \hat{a}^{\dag}\hat{a}\rangle$. When $\eta_{2} = 0.7173$ (red dashed line), the excitation number reaches its maximum value in the evolution process, which originates from the competition of RT and CRT. When $\eta_{2}$ reaches 1.2024, we can check that when $\tilde{g}_{r}$ approaches to zero, the CRT dominates the evolution. The higher excitation number of the resonator can be excited. The dynamics of the qubit and resonator show the periodic oscillation behavior. The results show that we can drive the system from JC regime to anti-JC regime through quantum Rabi regime (indicated by red dashed line).

\section*{Some applications on the quantum information theory}
\label{sec-IV}
Our scheme could be utilized as a candidate platform to implement the quantum information and computation device. As an example, we show the generations of Schr\"{o}dinger cat states and quantum gate. For this purpose, we first generalize our scheme to the multi-qubit case \cite{blais2007}. Considering $N$ qubits coupled to a resonator, we can obtain the simulated anisotropic quantum Dicke model with the same treatment. We assume all the qubits possess the same energy split ({\it i.e.}, $\varepsilon_{i}=\varepsilon$) and the periodic driving fields described in Eq.~(\ref{eq2c}) act on all the qubits. By means of the same approach, we can obtain the simulated anisotropic quantum Dicke model. The simulated anisotropic quantum Dicke model in the interaction picture reads
  \begin{eqnarray}
  \label{DM}
  \hat{H}_{\rm DM} &=& \tilde{g}_{r}\hat{a}\hat{J}_{+}e^{-i(\delta_{1}t+\varphi_{1})}+\tilde{g}_{cr}\hat{a}\hat{J}_{-}e^{-i(\delta_{2}t+\varphi_{2})}+{\rm H.c.},\nonumber\\
\end{eqnarray}
where $\hat{J}_{\pm}=\sum_{i}^{N}\hat{\sigma}_{i\pm}$ and $\hat{J}_{z}=\frac{1}{2}[\hat{J}_{+},\hat{J}_{-}]$. If we set the detunings of the blue and red sidebands $\delta_{1}=\delta_{2}=\delta$, we obtain the degenerate two-level system ({\it i.e.}, $\tilde{\varepsilon} = 0$), and the effective frequency of resonator is $\tilde{\omega}=\delta$. We also can adjust the normalized amplitudes of the driving fields to make $\tilde{g}_{r}=\tilde{g}_{cr}=\tilde{g}$. For simplicity, we set amplitudes $\eta_{1}=\eta_{2}$ and initial driving phases $\varphi_{i}=0$. In this case, the simulated Hamiltonian in the interaction picture reduces to the following form
\begin{eqnarray}
  \label{H_dis}
  \hat{H}_{\rm DM} &=&  \tilde{g}\left(\hat{a}^{\dag}e^{i\tilde{\omega} t}+\hat{a}e^{-i\tilde{\omega} t}\right)\hat{J}_{x}.
\end{eqnarray}
The evolution operator for the Hamiltonian in Eq.~(\ref{H_dis}), which could be obtained by means of the Magnus expansion, reads \cite{blanes2009}
\begin{eqnarray}
  \label{U_dis}
  \mathcal{U}(t) &=&  \exp[i\phi(t)\hat{J}_{x}^{2}]D[\xi(t)\hat{J}_{x}],
\end{eqnarray}
where $D(\xi)=\exp(\xi \hat{a}^{\dag}-\xi^{*}\hat{a})$, $\xi(t)=(\tilde{g}/\tilde{\omega})(1-e^{i\tilde{\omega}t})$ and $\phi(t)=(\tilde{g}/\tilde{\omega})^{2}(\tilde{\omega}t-\sin(\tilde{\omega}t))$.
Based on the dynamics of this effective Hamiltonian, the Schr\"{o}dinger cat states and quantum gate can be generated.

\subsection*{The generation of Schr\"{o}dinger cat states}
Superposition of coherent states plays an important role in quantum theory \cite{liao2016,huangjf2017,liu2005,liao2008,yin2013}. In this subsection, we consider how to generate superposition of coherent states for a single-qubit case. Assuming the initial state prepared on $\ket{\tilde{\psi}(0)}=\ket{0}_{r}\otimes\ket{g}$, we obtain the evolution state as follows
 \begin{eqnarray}
  \label{cat_dis1}
  \ket{\tilde{\psi}(t)}=\frac{e^{i\phi(t)}}{\sqrt{2}}(\ket{\xi(t)}_{r}\otimes\ket{+}-\ket{-\xi(t)}_{r}\otimes\ket{-}),
\end{eqnarray}
where $\ket{\pm}=\frac{1}{\sqrt{2}}(\ket{e}\pm\ket{g})$ are the eigenstates of $\hat{\sigma}_{x}$ and $\ket{\pm\xi(t)}_{r}=D[\xi(t)]\ket{0}_{r}$ are the coherent states with amplitude $\pm \xi(t)$. In the basis $\ket{e}$ and $\ket{g}$, the above state can be rewritten as following form
 \begin{equation}
  \ket{\tilde{\psi}(t)}=\frac{e^{i\phi(t)}}{2}(\ket{\mathcal{C}_{-}(t)}_{r}\otimes\ket{e}+\ket{\mathcal{C}_{+}(t)}_{r}\otimes\ket{g}),
\label{cat_dis2}
\end{equation}
where $\ket{\mathcal{C}_{\pm}(t)}=\mathcal{N}_{\pm}(\ket{\xi(t)}\pm \ket{-\xi(t)})$ with normalization coefficients $\mathcal{N}_{\pm}=\sqrt{2(1\pm \exp(-2|\xi(t)|^{2}))}$. Performing a projection measurement on the states $\ket{e}$ and $\ket{g}$, we obtain the states $\ket{\mathcal{C}_{+}(t)}$ and $\ket{\mathcal{C}_{-}(t)}$, which correspond to the even and odd Schr\"{o}dinger cat states. The magnitude of the displacement for $\ket{\mathcal{C}_{\pm}(t)}$ is $|\xi(t)|=2|(\tilde{g}/\tilde{\omega})\sin(\tilde{\omega}t/2)|$. When the evolution time $t_{0}=\pi/\tilde{\omega}$, the magnitude of the displacement reaches its maximum value $2|\tilde{g}/\tilde{\omega}|$.
\subsection*{The implementation of quantum gate}
In this subsection, we consider two-qubit case. Assuming the evolution time $T=2\pi/\tilde{\omega}$, we obtain $\xi(T)=0$ and $\phi(T)=2\pi(\tilde{g}/\tilde{\omega})^{2}$. The evolution operator is reduced to the following form
\begin{eqnarray}
  \label{U_gate}
  \mathcal{U}(T) &=&  \exp[i\phi(T)J_{x}^{2}].
\end{eqnarray}
where $J_{x}=\hat{\sigma}_{1x}+\hat{\sigma}_{2x}$ for two-qubit case. The Eq.~(\ref{U_gate}) can be rewritten as the form $\mathcal{U}(T)= \cos\vartheta \mathcal{I}+i \sin\vartheta \hat{\sigma}_{1x}\hat{\sigma}_{2x}$, where $\vartheta=2\phi(T)$ and $\mathcal{I}$ is identity operator for two-qubit. Here, we have omitted the total phase factor. To assess the capacity of the quantum gate, Zanardi {\it et. al.} introduced the entangling power \cite{zanardi1,zanardi2}. The entangling power for this unitary operator reads $e_{p}(\mathcal{U})=\frac{2}{9}\sin^{2}(2\vartheta)$. So the evolution operator can be viewed as a nontrival two-qubit quantum gate when $\theta\neq \frac{k}{2}\pi$ ($k$ is integer). When $\vartheta=\pi/4$ ({\it i.e.}, $\tilde{g}/\tilde{\omega}=0.25$), the quantum gate reads $\mathcal{U}(T)= \frac{1}{\sqrt{2}}\left( \mathcal{I}+i\hat{\sigma}_{1x}\hat{\sigma}_{2x}\right)$. Such quantum gate is local equivalent to the control-not (CNOT) gate \cite{Makhlin2002,Zhang2003}. The equivalent relation reads
\begin{eqnarray}
  \label{U_equi}
{\rm CNOT}=(u_{1}\otimes u_{2})\mathcal{U}(T)(u_{3}\otimes u_{4}),
\end{eqnarray}
where local unitary operators are as follows
\begin{equation}
\label{eq_local}
 \begin{array}{llll}
  u_{1}=\frac{1}{\sqrt{2}}\left(
                   \begin{array}{cc}
                     -1 & 1 \\
                     1 & 1\\
                   \end{array}\right),~u_{2}=\left(
                                  \begin{array}{cc}
                                    1 & 0 \\
                                    0 & 1 \\
                                  \end{array}
                                \right),~u_{3}= \frac{1}{\sqrt{2}}\left(
  \begin{array}{cc}
     -1 & -i \\
    1 & -i \\
  \end{array}\right),~ u_{4}=\frac{1}{\sqrt{2}}\left(
                  \begin{array}{cc}
                  1 & i \\
                  i & 1 \\
                  \end{array}
                   \right).
\end{array}
\end{equation}

\section*{Discussion}
In conclusion, we have proposed a method to simulate a tunable AQRM, which is achieved by driving the qubit(s) with two-tone periodic driving fields. We have analyzed the parameter conditions under which this proposal works well. By choosing proper modulation frequencies and amplitudes, the coupling constants of RT or CRT can be suppressed to zero, respectively. Consequently, we study the dynamics induced by CRT or RT correspondingly. In addition, we have also discussed the applications of our scheme to the generations of quantum gate and Schr\"{o}dinger cat states. This proposal provides us with a reliable approach for studying the effects of RT and CRT in different regimes individually. Although we explore the scheme with the circuit QED system, which could be implemented in other systems, {\rm e.g.}, cavity QED and trapped ion systems. The presented proposal will pave a way to further study the stronger light-matter interaction in a system whose coupling strength is far away from the USC and DSC regimes in quantum optics.

Extensions of presented scheme to a variety of physically relevant systems, such as multi-qubit and multi-mode fields interaction system and the system coupling with the environments, deserve future investigations.


\section*{Acknowledgements}
The work is supported by the NSF of China (Grant No. 11575042).

\section*{Author Contributions}
G.W., H.S., and K.X. initiated the idea. C.S. and R.X. developed the model and performed the calculations. G.W. provided numerical results. All authors developed the scheme and wrote the main manuscript text.

\vspace{2pt}

\section*{Additional Information}
\textbf{Competing Interests:} The authors declare no competing interests.
\end{document}